\documentclass[journal, final, 10pt, twocolumn]{IEEEtran}

\usepackage[letterpaper, left=.92in, right=.92in, bottom=.84in, top=.75in]{geometry}
\usepackage{microtype}
\usepackage{caption}
\usepackage{graphicx}
\usepackage{amsmath}

\usepackage{array}
\usepackage{amssymb}
\usepackage{cite}
\usepackage{url}

\usepackage{xcolor}

\usepackage[ruled,lined,linesnumbered]{algorithm2e}

\usepackage{caption}
\usepackage{color,soul}
\usepackage[T1]{fontenc}
\usepackage[utf8]{inputenc}

\begin{document}

\title{Decentralized Energy Marketplace via NFTs and AI-based Agents}

\author{
\IEEEauthorblockN{Rasoul Nikbakht},
\IEEEauthorblockN{Farhana Javed},
\IEEEauthorblockN{Farhad Rezazadeh},
\IEEEauthorblockN{Nikolaos Bartzoudis},
\IEEEauthorblockN{Josep Mangues-Bafalluy}\\

Centre Tecnològic Telecomunicacions Catalunya (CTTC/CERCA), 08860 Castelldefels, Spain\\
Email: \{rasoul.nikbakht,  farhana.javed, farhad.rezazadeh, nikolaos.bartzoudis, josep.mangues\}@cttc.es
}

\maketitle
\thispagestyle{empty}
\begin{abstract}

The paper introduces an advanced Decentralized Energy Marketplace (DEM) integrating blockchain technology and artificial intelligence to manage energy exchanges among smart homes with energy storage systems. The proposed framework uses Non-Fungible Tokens (NFTs) to represent unique energy profiles in a transparent and secure trading environment. Leveraging Federated Deep Reinforcement Learning (FDRL), the system promotes collaborative and adaptive energy management strategies, maintaining user privacy. A notable innovation is the use of smart contracts, ensuring high efficiency and integrity in energy transactions. Extensive evaluations demonstrate the system's scalability and the effectiveness of the FDRL method in optimizing energy distribution. This research significantly contributes to developing sophisticated decentralized smart grid infrastructures. Our approach broadens potential blockchain and AI applications in sustainable energy systems and addresses incentive alignment and transparency challenges in traditional energy trading mechanisms. The implementation of this paper is publicly accessible\footnote{Available at \url{https://github.com/RasoulNik/DEM}}.
\end{abstract}

\section{Introduction}

In the evolving landscape of energy distribution, there is an emergent need to manage excess energy in a manner that is both decentralized and secure. Traditional paradigms, anchored in centralized architectures, face challenges in transparency, operational efficiency, and incentive misalignment, focusing more on grid operator benefit rather than the mutual interests of the grid operator and households. As we steer towards a more sustainable future, modern households are transforming from mere energy consumers into dual-role entities — both consumers and producers. However, to facilitate this transition, we need to establish a system that incentivizes both users and the grid to participate in fair energy trade.

The Decentralized Energy Marketplace (DEM), proposed in this work in based on the blockchain technology and introduces a smart contract-based framework for energy trading. DEM utilizes three core components: an \textit{Energy Profile} represented by Non-Fungible Tokens (NFTs), an \textit{Energy Pool}, and a \textit{Market Contract}. These contracts facilitate the trading, validation, and settlement processes in the energy market, with further details provided later.

In our system, modern households equipped with batteries and AI-based agents form an advanced grid. These agents, capable of collaborating through federated learning (FL), enable decentralized model training across multiple households without data centralization. This collaboration aids in trusted, real-time energy distribution, responding to demand and supply dynamics. The integration of DEM, household batteries, and AI agents constitutes what we term a \textit{Virtual Power Plant} (VPP).

Smart contracts ensure transparency and adds a level of trust in our system but face challenges in executing data-heavy operations like AI or complex data analyses on the blockchain, mainly due to high computational costs. The Zero-Knowledge Machine Learning (ZK-ML) stack, as discussed in \cite{xing2023zero}, addresses this by allowing AI agents to function off-chain, balancing security, privacy, and computational efficiency. This approach also permits offloading other computational tasks of the trading platform to off-chain entities. Our paper details the system's conceptual framework and mechanics but only touches on integrating agents with ZK-ML, reserving a thorough analysis for future research.



In our decentralized energy trading architecture, NFTs transcend their role as mere digital artifacts to become computational entities encapsulating user identity and contractual commitments. Users from households to grid operators create EnergyProfile NFTs that hold key data and manage collateral, ensuring protocol compliance. These dynamic NFTs record energy commitments, aggregating in the EnergyPool smart contract for a unified supply-demand system. This approach, extending beyond traditional NFT uses, necessitates blockchain oracles for enhanced platform validation.

Blockchain oracles, notably Chainlink \cite{breidenbach2021chainlink}, are crucial in linking the deterministic blockchain systems with fluctuating real-world data. They can be centralized or decentralized, using various consensus mechanisms to ensure data integrity. Our architecture uses Chainlink’s decentralized oracle network to infuse smart contracts with up-to-date energy and pricing data.

The primary innovative contributions of this paper can be summarized as follows:

\begin{itemize}
  \item \textbf{Utilizing Blockchain technology:} Incorporating NFTs to represent individual energy profiles on the blockchain, providing a novel method for managing and trading energy in a smart contract-based marketplace.
  \item \textbf{Advancing AI in Energy Systems:} Employing FDRL to foster collaborative and private energy management across households, enhancing the traditional energy trading models.
  \item \textbf{Decentralized VPP:} Merging state-of-the-art AI agents with household batteries to create a Virtual Power Plant.
  \item \textbf{Oracle integration:} Utilizing decentralized oracles, specifically Chainlink, to securely and reliably integrate real-world (price and energy) data with smart contracts for comprehensive and fair energy trading.
\end{itemize}
Finally, the proposed DEM framework, in conjunction with the AI agent, transforms the surplus energy of households into a reliable energy source, functioning as a Virtual Power Plant (VPP). In this scenario, the main grid sets the energy price and can incentivize users to actively participate in energy production by adjusting the price, thereby leading to mutual benefit and resolving the incentive misalignment problem.

\section{Related work}

Several studies have investigated Local Energy Market (LEM) designs. Zhou et al. \cite{zhou2023peer} focus on peer-to-peer energy sharing, emphasizing AI, machine learning, and IoT's role in clean energy transition, with blockchain enhancing security and efficiency. Other research includes centralized and decentralized strategies in high PV contexts \cite{Community-Based-Energy-Market2021}, Distributed Ledger Technology (DLT) for energy trading \cite{Distributed-Ledger-Based-AMM-smart-grid2022}, and blockchain for virtual power plants \cite{yang2021blockchain}, featuring a privacy-centric algorithm. Lee et al. \cite{lee2023blockchain} discuss smart grid advancements with AI and blockchain, while Yu et al. \cite{yu2023distributed} combine these technologies for optimized power trading, addressing scalability and communication efficiency. Chien et al. \cite{chien2023prediction} propose a blockchain-based peer-to-peer market with an LSTM neural network for improved energy consumption and production predictions.

Our contribution is distinguished from the existing literature, which focuses on blockchain and DLT, by proposing the novel use of NFTs to represent user identities and energy commitments, offering a unique approach to decentralized energy trading. We leverage blockchain oracles, specifically Chainlink, to bridge real-world data with the blockchain, and introduce a framework utilizing NFTs alongside smart contracts, thus expanding the blockchain's application in smart grids and setting a new paradigm for decentralized energy exchanges.

\begin{figure}[t]
    \centering
    \includegraphics[width=0.45\textwidth]{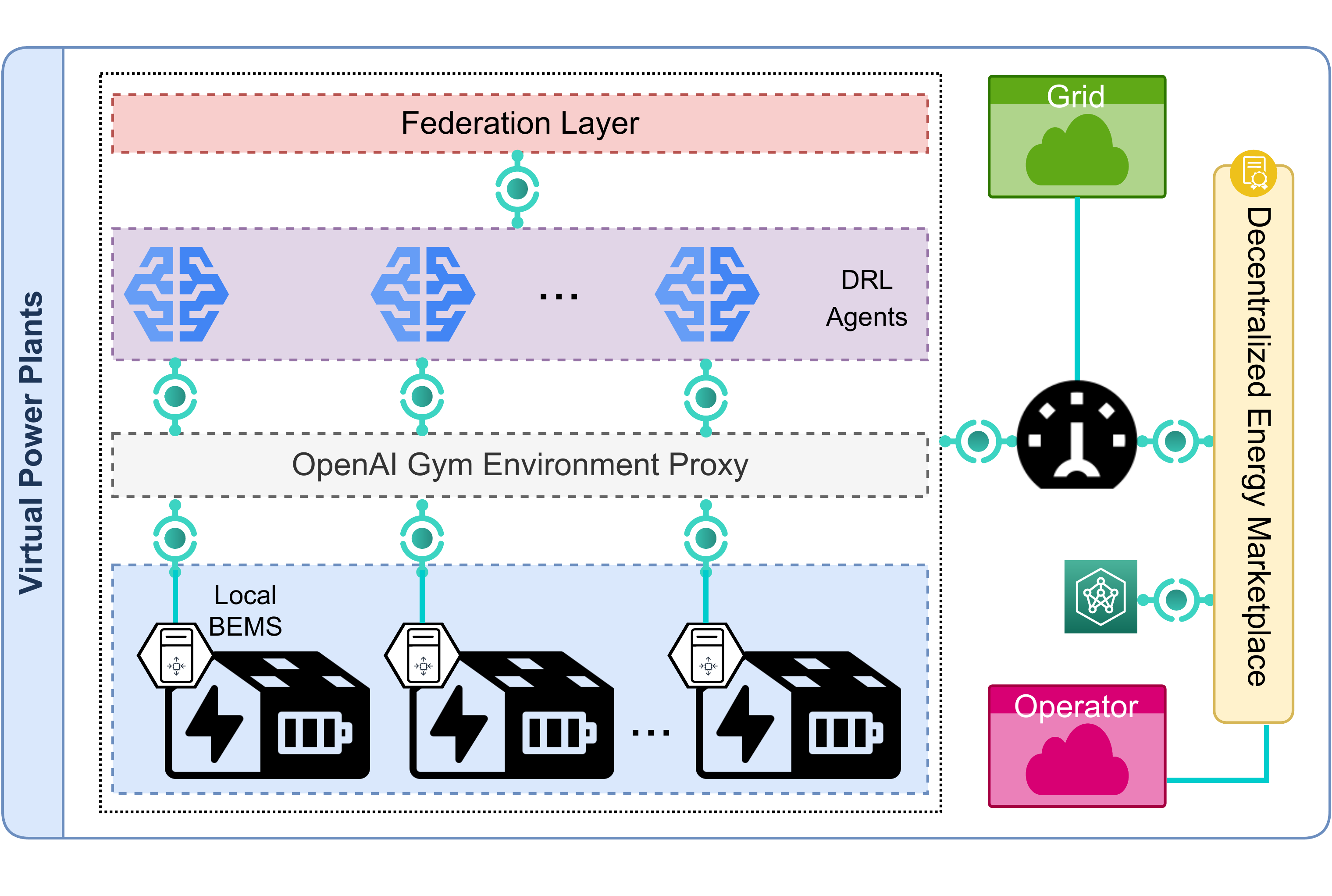}
    \caption{Proposed System Model.}
    \label{fig:SystemModel}
    \vspace{-.5cm}
\end{figure}

\section{System Model}
This section explains the proposed system model, as depicted in Figure \ref{fig:SystemModel}. Building upon the concepts from our prior work \cite{rezazadeh2022federated}, here our proposed Distributed Energy Marketplace (DEM) employs a Federated Deep Reinforcement Learning (FDRL) approach. The system facilitates energy management across various smart buildings by allowing them to not only train their models but also to collaboratively share insights. This synergy promises enhanced efficiency, cost reduction, and a decrease in CO2 emissions, especially in smart grids integrated with PV systems and batteries.

The primary components of the system model, as outlined in Fig \ref{fig:SystemModel}, are:
\begin{itemize}
    \item \textbf{Smart Homes with Energy Storage:} These homes are outfitted with battery storage and leverage AI to monitor and manage energy use. 
    \item \textbf{Decentralized Energy Marketplace (DEM):} This platform is a decentralized application (DApp) that ensures the secure and transparent exchange of energy, orchestrating excess energy production in a decentralized manner. 
    \item \textbf{Federated Deep Reinforcement Learning (FDRL) Framework:} By combining FL and Deep Reinforcement Learning, this framework enables smart buildings to collectively refine their energy management strategies. It ensures enhanced decision-making for the grid and each smart home, all while maintaining data privacy.
    \item \textbf{Virtual Power Plants and Grid Operators:} Through the integration of the DEM, smart homes with energy production and consumption capabilities, and the  FDRL Agents that act in the behavior of each smart home, a Virtual Power Plant (VPP) is created. This VPP manages the surplus energy from smart homes, creating a reliable energy source for the main energy grid.
\end{itemize}

This setup shown in Fig \ref{fig:SystemModel} is designed to meet the power grid's needs efficiently and safely. However, this innovative approach raises important questions about how AI agents are used, how trustworthy they are in the DEM, and how to make them reliable. A key answer to these concerns is the ZK-ML method \cite{xing2023zero}, which guarantees that the system's parts run the algorithms correctly.

ZK-ML principles complement Federated Learning (FL), fostering decentralized and collaborative model training without explicit data exchange. This approach ensures uniform execution and robust threat protection. A notable implementation is the ezkl library, which converts TensorFlow or PyTorch computational graphs into zero-knowledge proofs (ZK-SNARK circuits), allowing secure verification of operations by FDRL agents on private data. This paper briefly mentions ZK-ML's future potential without detailed exploration.

Our FDRL framework merges FL with Deep Reinforcement Learning (DRL) for improved decentralized energy management. In this setup, households equipped with batteries and solar panels serve as nodes in a micro-grid Energy Management System (EMS). Local agents at these nodes train DRL models and share hyperparameters for collective improvement. This intelligence is integrated within a federation layer, forming a global model in the EMS for autonomous decision-making. Our approach envisions a network of Building Energy Management Systems (BEMS), where each household contributes data, creating a decentralized and collaborative environment for optimal energy management while maintaining privacy and integrity.

\begin{figure}
    \centering
    \includegraphics[width=.4\textwidth]{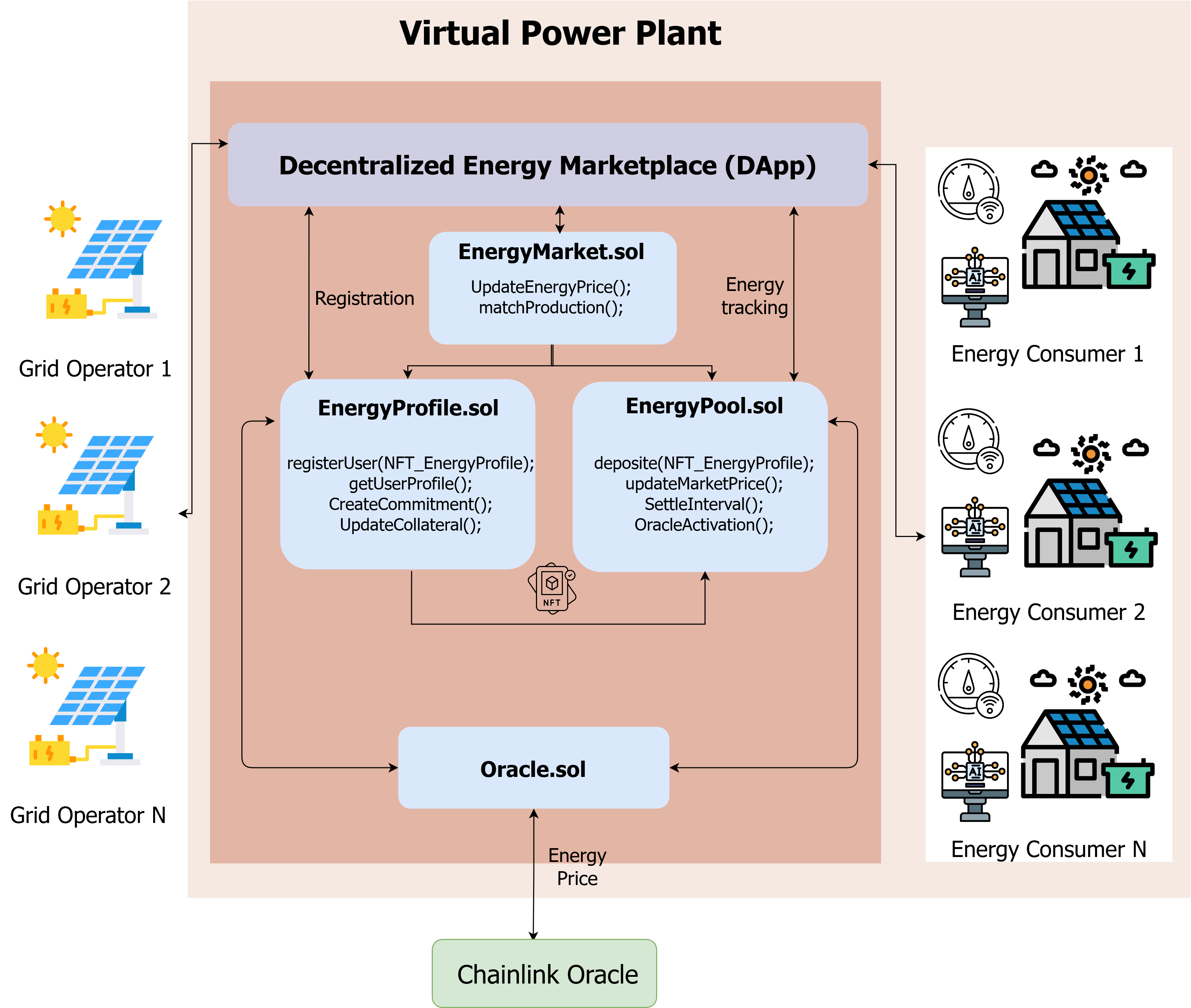}
    \caption{User interaction with DEM and its key smart contracts.} 
    \label{fig:Components_Interactions}
    \vspace{-.5cm}
\end{figure}

\section{Blockchain-enabled Smart Decentralized Energy Marketplace}
Our solution integrates the concept of NFTs to empower small energy producers, granting them unparalleled autonomy over their energy generation and facilitating their active participation in the \textit{decentralized energy marketplace}. 

We divide this section into two parts: firstly, we explain the proposed blockchain components and their interaction. Secondly, we elucidate the AI-based agents system and algorithms.

\subsection{Proposed Blockchain-enabled smart contracts and key components}

Fig. \ref{fig:Components_Interactions} illustrates the components of the DEM (smart contracts and chainlink Oracle) and their interactions. To provide an in-depth understanding, we explain the key operations of each smart contract below:

\begin{itemize}
    \item \textbf{EnergyProfile Contract:} 
    This contract manages the energy profiles of participants. Each participant has a unique NFT representing their energy attributes, including commitments for consumption or production, and the fulfillment of these commitments. It also enables DEM users to create energy commitments for consumption and production.
    
    \item \textbf{EnergyPool Contract:} 
    This contract handles the energy commitments from users once their profile is deposited into the contract. It tracks the total committed energy for production and consumption and labels the commitments as processed after they pass through this contract.
    
    \item \textbf{EnergyMarket Contract:} 
    This contract serves as the operational core, linking the EnergyProfile and EnergyPool contracts. It manages commitment transactions and temporarily stores them in a MarketBuffer until they are completed. It also settles payments using the price and energy Oracle.
\end{itemize}

\begin{figure}[t]
    \centering
    \includegraphics[width=0.45\textwidth]{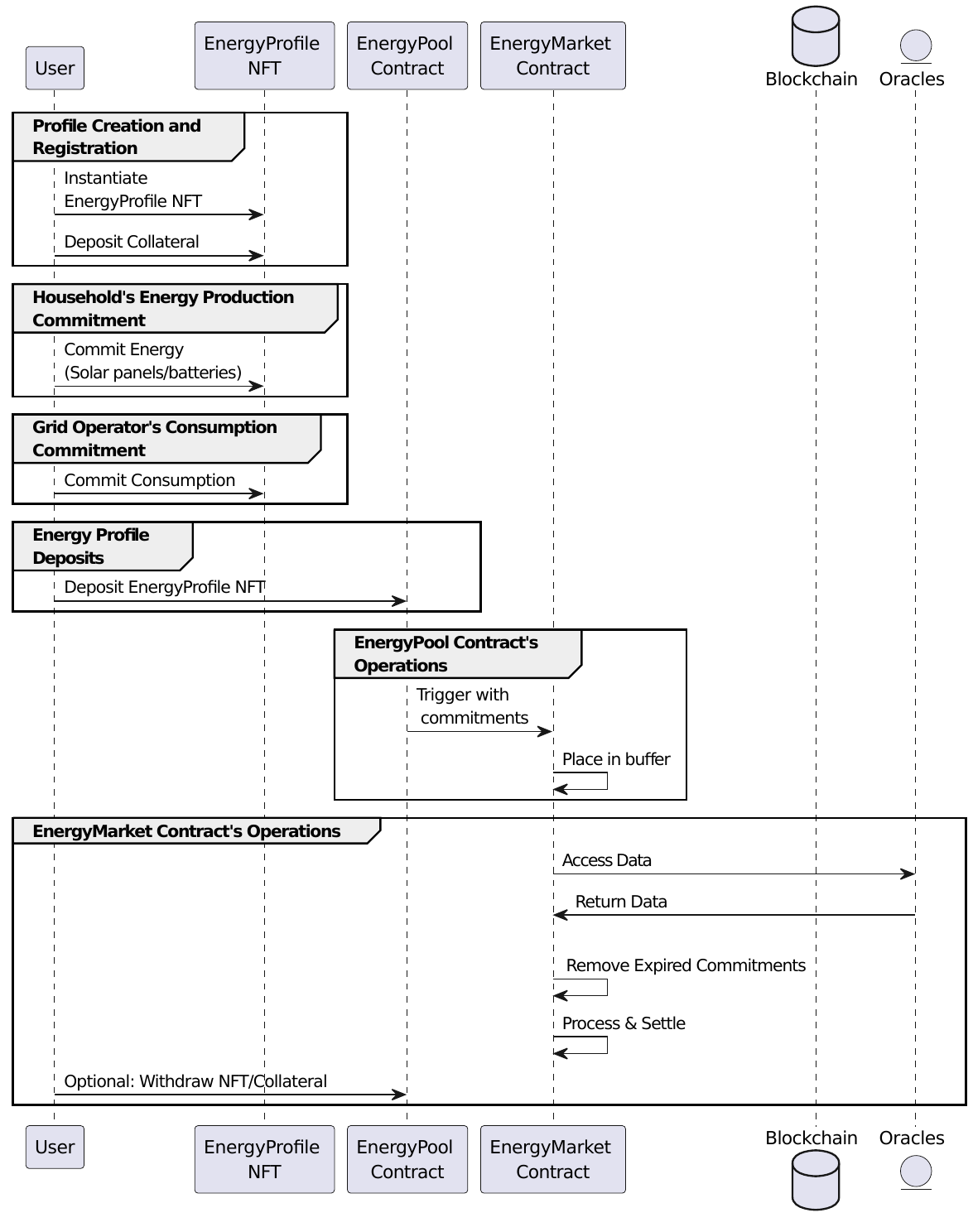}
    \caption{Smart contracts flow and interaction sequence.}
    \label{fig:sequence}

\end{figure}

Fig. \ref{fig:sequence} illustrates the interactions among users, smart contracts, and the blockchain within the decentralized energy marketplace. It highlights the process of creating, updating, and transacting energy profiles through NFTs, followed by the processing of commitments and settlements. The following actions are executed for the end-to-end operation of the DEM:
\begin{enumerate}
    \item \textbf{Profile Creation and Registration:} Users, encompassing both households and grid operators, instantiate their EnergyProfile NFTs. These tokens encapsulate pivotal user data and proffer a suite of functionalities, notably depositing and withdrawing user collateral. The collateral mechanism serves as a deterrent, ensuring adherence to established protocol stipulations; non-compliance results in collateral forfeiture.
    
    \item \textbf{Household's Energy Production Commitment:} Households (represented as AI agents) make energy production commitments by specifying the amount of energy they can deliver. This energy can come from solar panels or batteries and is determined for the upcoming time slots (each time slots is 1 hour). Agents decide the amount of energy and its source (either batteries or solar panels) based on household consumption patterns.
    
    \item \textbf{Grid Operator's Consumption Commitment:} Grid operators, spanning both primary and virtual entities, proffer consumption commitments aligned with forthcoming time slots.
    
    \item \textbf{Energy Profile Deposits:} All users, including households and grid operators, deposit their EnergyProfile NFTs, which contain their commitments, into the EnergyPool contract.
    
    
    \item \textbf{EnergyPool Contract's Operations:} The EnergyPool contract aggregates supply and demand, triggering the EnergyMarket contract upon the arrival of new commitments. It places these commitments into the temporary buffer of the EnergyMarket contract. (For now, we assume a single large consumer akin to a virtual power plant; this important assumption and the virtual power plant concept must be introduced.)
    
    \item \textbf{EnergyMarket Contract's Operations:} The EnergyMarket contract removes expired commitments from the buffer, processes, and settles the commitments. Users can then withdraw their NFTs or their collateral if they wish. The EnergyMarket contract also has access to price and energy oracles that provide off-chain information to the contract.
\end{enumerate}

The system allows users to maintain a maximum of three concurrent commitments. Users can substitute processed commitments in their profile with new ones without having to withdraw their EnergyProfile NFT from the EnergyPool contract. Therefore, users can continue interacting with the DEM by updating their energy commitments.

\subsection{AI-based Agents: Proposed FDRL Framework and local training}
Building on the principles of collaborative learning and energy optimization, our proposed FDRL (Federated Deep Reinforcement Learning) framework establishes the basis for a decentralized microgrid management system. Within this system, each household acts as a node, equipped with a solar panel and storage unit. These nodes collectively operate in designated time slots, making decisions about energy production and consumption. We compare our approach with local Deep Reinforcement Learning (DRL) agents. Both local and federated approaches aim to minimize external energy usage (weighted with the respective energy price in each time slot of system operation) and reduce household energy bills. 

\subsubsection{Proposed FDRL Framework}
\label{sec:scenario}

We define a micro-grid with a single federation layer (located in the EMS) and a set of houses, \( \mathcal{H} \), each equipped with a Photovoltaic (PV) system and a battery \( \mathcal{B} \) of capacity \( h_c \). The system operates in time slots, divided into decision intervals \( t \in \mathcal{T} = \{1, 2, \ldots, T\} \), with actions initiated at each interval's start. The federation layer collects and shares models every \( \hat{T} \) intervals, termed as the federation episode. For global federated model formation, various strategies can be employed. Given \( \omega_{h} \) as the weight of a local DRL agent in EBMS, the aggregated global model for \( \mathcal{H} \) EBMS is:
\[
\omega_G = \frac{1}{|\mathcal{H}|} \sum_{h=1}^{|\mathcal{H}|} \omega_h
\]

Our approach's advantages include: 
\begin{enumerate}
    \item Enhanced power control in the microgrid, ensuring timely and precise data usage. 
    \item Significant reduction in control information traversing the microgrid, mitigating EMS overhead and potential bottlenecks.
    \item Information sharing among local BEMS, leveraging FL to boost DRL agent capabilities by refining the learning process's generalization.
\end{enumerate}
The details of the FDRL are presented in Algorithm \ref{algo:FDRL}.
We aim to minimize the cost of external energy input and household energy bills:
\begin{equation}
\min \sum_{t=1}^{T} E\left[\sum_{h\in H} C_h^{(t)}\right],
\end{equation}
where \(C_h^{(t)}\) represents the cost of external energy supplied minus energy profit that is delivered to the main grid considering in the given time slot $t$. This optimization problem is cast as a Markov Decision Process (MDP) to household energy bills (in some cases it might lead to net profit). The agents, guided by AI, learn to charge or discharge batteries and trade energy based on various conditions. We employ a reward-penalty method to align the MDP's reward setting with our models. For actions, the local DRL agents (BEMS) consider:

1) Trading energy surpluses and deficits with an external network, bypassing the battery (\(a_1\)), 2) Charging the battery using surplus electricity (\(a_2\)), and 3) Discharging the battery to compensate for energy consumption exceeding production (\(a_3\)). Thus, the action space is \(\mathcal{A} = \{a_1, a_2, a_3\}\). Our FDRL framework seeks long-term energy cost optimization by selecting the most appropriate action for varying micro-grid states. The state space comprises: 1) Energy from the PV system (\(s_1\)), 2) Current battery capacity (\(s_2\)), 3) Daytime temperature (\(s_3\)), and 4) Power consumption (\(s_4\)). Formally, the state space is given by:
\begin{equation}
s_i^{(t)} = \{ ( s_{1,h}^{(t)}, s_{2,h}^{(t)}, s_{3,h}^{(t)}, s_{4,h}^{(t)} ) \mid \forall h \in \mathcal{H} \}
\end{equation}
generated/stored energy and external supply. It is defined for the hh-th agent as:
\begin{equation}
r^{(t)}_h = C_h^{(t)} - \alpha \times \text{relu}(-\mathcal{B} + 0.1 \times h_c)
\end{equation}
In this equation, the relurelu function, outputs its input if the input is positive and zero otherwise, acting as a threshold mechanism. The coefficient $\alpha$ regulates the impact of the second term, which is designed to maintain a healthy battery level above $10\%$. This term penalizes scenarios where the battery level falls below this threshold, with $\alpha$ adjusting the severity of this penalty.
\subsubsection{Local Training Algorithm}
We use discretized soft actor-critic (SAC)~\cite{DiSAC, rezazadeh2022federated}, an actor-critic technique, that integrates policy optimization and Q-Learning as a benchmark for local training. For a micro-grid environment, we define $\rho_\pi(s_t)$ and $\rho_\pi(s_t, a_t)$ as the state and state-action distributions induced by policy $\pi(a_t|s_t)$ respectively. In contrast to deep deterministic policy gradient (DDPG) \cite{DDPG}, SAC employs a stochastic policy gradient \cite{cont4}. Policy-based algorithms aim to modify the policy parameters $\phi$ in line with the performance gradient $\nabla_{\phi}J(\pi_{\phi})$ as per the policy gradient theorem \cite{cont3}. 

\begin{algorithm}[t]
\scriptsize
\SetKwInOut{Input}{Input}
\SetKwInOut{Output}{Output}
\SetKwInOut{Return}{return}
\Input{$t, T, \omega_{h}^{(t)}$ $\forall h \in \mathcal{H}$}
\Output{Improved federation models $\omega_{G}^{(t+1)}$}
 
  \If{$mod(t,\hat{T})==0 \land t > 0$}{
          \For{ each $\omega_{h}^{(t)}$ $\forall h \in \mathcal{H}$, in parallel}{
                Collect $\omega_{h}^{(t)}$\;
                $\omega_{G}^{(t+1)} \leftarrow \frac{1}{\mathcal{H}} \sum_{h=1}^{\mathcal{H}}\omega_h$\;
                $\omega_{h}^{(t+1)} \leftarrow \omega_{G}^{(t+1)}$\;
         }
        \emph{ \#Return updated local models}\;  \
        \Return{$\omega_{h}^{(t+1)}, \forall h \in \mathcal{H}$}
    }
    Run SAC algorithm~\cite{DiSAC, rezazadeh2022federated}\;

\caption{Weight update process in federation layer}
\label{algo:FDRL}
\end{algorithm}
\section{Performance Evaluation}
The performance evaluation of the proposed DEM and FDRL is presented in the following section. Integrating these two modules is relatively straightforward as they operate independently. Therefore, we assess their performance separately, examining their effectiveness in two independent scenarios.

\subsection{Deployment and Testing with Brownie and Ganache} 
The proposed DEM framework offers a 
solution for the management and exchange of energy profiles on Ethereum Virtual Machine (EVM)-compatible blockchains. Deployment and testing of the DEM are conducted using the Brownie development environment and the Ganache blockchain simulator. Furthermore, mock oracle contracts from Chainlink (MockV3Aggregator) are employed to simulate energy production and price data, ensuring the DEM system's operation with comprehensive components.

For the deployment of the DEM system, the following steps are undertaken:
\begin{itemize}
    \item The Ganache blockchain simulator is initiated with 50 accounts.
    \item The DEM contract, along with the Chainlink mock contracts, is deployed.
    \item An end-to-end DEM framework with a varying number of users is simulated.
\end{itemize}

\begin{figure}
    \centering
    \includegraphics[width=.4\textwidth]{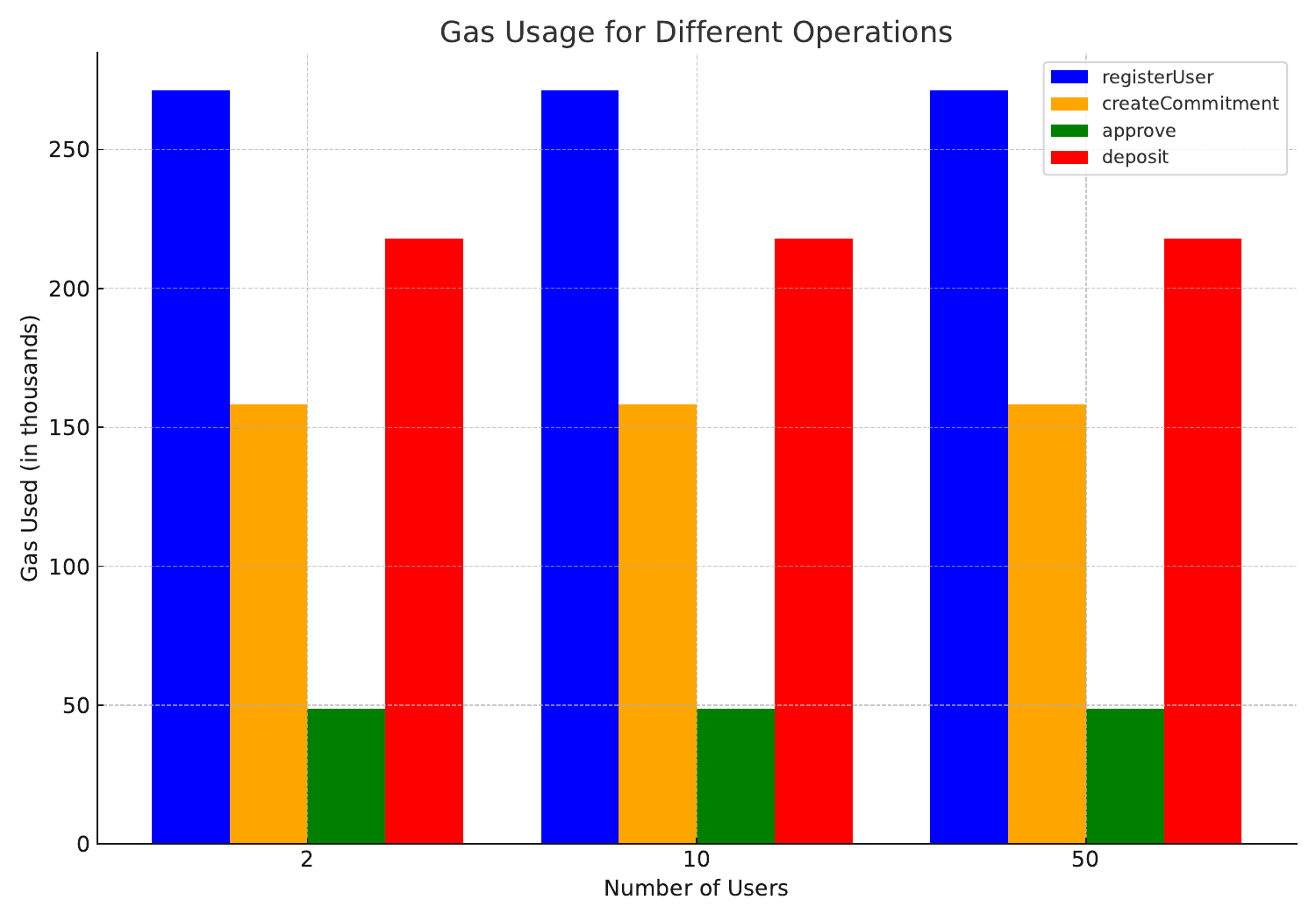}    
    \caption{Gas used in blockchain for different interactions with smart contracts with the number of users increasing from 2 to 50. } 
    \label{fig:gas-usage}
\end{figure}

In our end-to-end tests, we simulate environments with 2, 10, and 50 households, in addition to a main grid. Users register within the system by minting Profile NFTs. Households commit to energy production, and the main grid commits to energy consumption. Gas consumption for transactions is measured and reported for key functions of the system. Figure~\ref{fig:gas-usage} illustrates the gas consumption for each operation within the DEM. Notably, gas usage, which reflects the transaction costs on the Ethereum blockchain, stabilizes after a marginal decline from 2 to 10 users, demonstrating the framework's efficiency and scalability. Moreover, the gas usage for each operation remains significantly below the Ethereum block's average gas limit of 30 million, reinforcing the practicality of our framework. We can greatly reduce transaction costs by deploying our framework to Ethereum Layer 2 solutions like Arbitrum, on EVM-based public chains such as Polygon, or on EVM-compatible private chains including Hyperledger projects.

\subsection{Training of AI Agents}
Our DRL architecture is built in Python, incorporating the OpenAI Gym library~\cite{Globe_far}. Furthermore, each DRL-agent is equipped with a SAC algorithm, as detailed in~\cite{DiSAC}. To train the DRL agent, we utilize a dataset\footnote{https://github.com/antoine-delaunay/DRL\_SmartGrid/tree/main/Data} comprising actual household measurements from July 15, 2016, to July 15, 2019. This dataset includes input features like air temperature, total electricity consumption, and energy generated by the PV system, recorded every five minutes.
To minimize transitions on the blockchain, we restricted the decision interval for the AI agents to one-hour slots, in line with our DEM deployment strategy. We derive results for four households and report the average performance. 
\begin{figure}
    \centering
    \includegraphics[width=.45\textwidth]{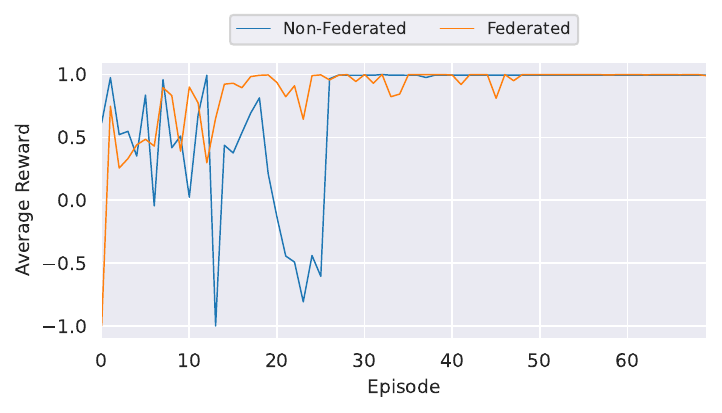}    
    \caption{Comparison of federation versus non-federation approaches, illustrating the average performance across four households.} 
    \label{fig:fed-nonfed}
\end{figure}

\begin{figure}[t!]
    \centering
    \includegraphics[width=.45\textwidth]{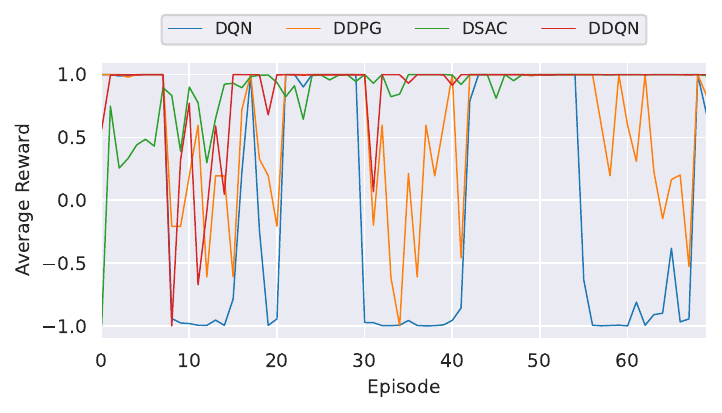}    
    \caption{Average reward for the first house across different approaches.} 
    \label{fig:feds}
\end{figure}

In Figure \ref{fig:fed-nonfed}, the experimental results clearly demonstrate a higher and more stable average reward in the federation approach as compared to the non-federation approach based on SAC algorithm.
This suggests that the collaborative learning environment in the federation model significantly contributes to a more robust and efficient convergence.
Figure \ref{fig:feds} illustrates the performance of various reinforcement learning techniques in a federated setting, where we observe that SAC surpasses other algorithms with the given dataset and reward function. Indeed, the descritized SAC is capable of learning optimal actions and strategies for managing energy over a specific period.  

\section{Conclusion}
This paper introduces a pioneering Decentralized Energy Marketplace (DEM) that merges blockchain and artificial intelligence, enabling sustainable energy transactions for smart homes. NFTs uniquely represent energy profiles, and Federated Deep Reinforcement Learning (FDRL) enhances energy management collectively while safeguarding privacy. The framework's foundation on smart contracts ensures that energy trades are secure and transparent. Computational efficiency is showcased through reduced agent convergence times and scalability validation. Looking ahead, future work will focus on integrating off-chain AI agent execution via the Zero-Knowledge Machine Learning (ZK-ML) stack, aiming to fortify security and reduce blockchain computational demands. This evolution is expected to uphold data privacy and boost transaction cost efficiency, further advancing the potential of decentralized smart grid infrastructures.
\section*{Acknowledgment}
This work was partially funded by the Spanish Government (MICCIN \& NextGenEU program) under Grant PCI2020-112049, by the ECSEL Joint Undertaking (JU) under grant agreement No 876868, and by the Catalan government under the 2021 SGR 00772 grant. Additional funding was provided by the Generalitat de Catalunya under grant 2021 SGR 00770 (6GE2E) and by the MCIN/AEI/10.13039/501100011033 under grant PID2021-126431OB-I00 (ANEMONE).
\bibliographystyle{IEEEtran}
\bibliography{DEM}

\begin{thebibliography}{10}
\providecommand{\url}[1]{#1}
\csname url@samestyle\endcsname
\providecommand{\newblock}{\relax}
\providecommand{\bibinfo}[2]{#2}
\providecommand{\BIBentrySTDinterwordspacing}{\spaceskip=0pt\relax}
\providecommand{\BIBentryALTinterwordstretchfactor}{4}
\providecommand{\BIBentryALTinterwordspacing}{\spaceskip=\fontdimen2\font plus
\BIBentryALTinterwordstretchfactor\fontdimen3\font minus
  \fontdimen4\font\relax}
\providecommand{\BIBforeignlanguage}[2]{{%
\expandafter\ifx\csname l@#1\endcsname\relax
\typeout{** WARNING: IEEEtran.bst: No hyphenation pattern has been}%
\typeout{** loaded for the language `#1'. Using the pattern for}%
\typeout{** the default language instead.}%
\else
\language=\csname l@#1\endcsname
\fi
#2}}
\providecommand{\BIBdecl}{\relax}
\BIBdecl

\bibitem{xing2023zero}
Z.~Xing, Z.~Zhang, J.~Liu, Z.~Zhang, M.~Li, L.~Zhu, and G.~Russello,
  ``Zero-knowledge proof meets machine learning in verifiability: A survey,''
  \emph{arXiv preprint arXiv:2310.14848}, 2023.

\bibitem{breidenbach2021chainlink}
L.~Breidenbach, C.~Cachin, B.~Chan, A.~Coventry, S.~Ellis, A.~Juels,
  F.~Koushanfar, A.~Miller, B.~Magauran, D.~Moroz \emph{et~al.}, ``Chainlink
  2.0: Next steps in the evolution of decentralized oracle networks,''
  \emph{Chainlink Labs}, vol.~1, pp. 1--136, 2021.

\bibitem{zhou2023peer}
Y.~Zhou and P.~D. Lund, ``Peer-to-peer energy sharing and trading of renewable
  energy in smart communities-- trading pricing models, decision-making and
  agent-based collaboration,'' \emph{Renewable Energy}, 2023.

\bibitem{Community-Based-Energy-Market2021}
J.~L. Crespo-Vazquez, T.~AlSkaif, A.~M. Gonzalez-Rueda, and M.~Gibescu, ``A
  community-based energy market design using decentralized decision-making
  under uncertainty,'' \emph{IEEE Transactions on Smart Grid}, vol.~12, no.~2,
  pp. 1782--1793, 2021.

\bibitem{Distributed-Ledger-Based-AMM-smart-grid2022}
\BIBentryALTinterwordspacing
D.~B. Gajić, V.~B. Petrović, N.~Horvat, D.~Dragan, A.~Stanisavljević,
  V.~Katić, and J.~Popović, ``A distributed ledger-based automated
  marketplace for the decentralized trading of renewable energy in smart
  grids,'' \emph{Energies}, vol.~15, no.~6, 2022. [Online]. Available:
  \url{https://www.mdpi.com/1996-1073/15/6/2121}
\BIBentrySTDinterwordspacing

\bibitem{yang2021blockchain}
Q.~Yang, H.~Wang, T.~Wang, S.~Zhang, X.~Wu, and H.~Wang, ``Blockchain-based
  decentralized energy management platform for residential distributed energy
  resources in a virtual power plant,'' \emph{Applied Energy}, vol. 294, p.
  117026, 2021.

\bibitem{lee2023blockchain}
C.-D. Lee, J.-H. Li, and T.-H. Chen, ``A blockchain-enabled authentication and
  conserved data aggregation scheme for secure smart grids,'' \emph{IEEE
  Access}, 2023.

\bibitem{yu2023distributed}
Y.~Yu, J.~Wu, G.~Li, and W.~Wang, ``A distributed power trading scheme based on
  blockchain and artificial intelligence in smart grids.'' \emph{Intelligent
  Automation \& Soft Computing}, vol.~37, no.~1, 2023.

\bibitem{chien2023prediction}
I.~Chien, P.~Karthikeyan, and P.-A. Hsiung, ``Prediction-based peer-to-peer
  energy transaction market design for smart grids,'' \emph{Engineering
  Applications of Artificial Intelligence}, vol. 126, p. 107190, 2023.

\bibitem{rezazadeh2022federated}
F.~Rezazadeh and N.~Bartzoudis, ``A federated drl approach for smart micro-grid
  energy control with distributed energy resources,'' in \emph{2022 IEEE 27th
  International Workshop on Computer Aided Modeling and Design of Communication
  Links and Networks (CAMAD)}.\hskip 1em plus 0.5em minus 0.4em\relax IEEE,
  2022, pp. 108--114.

\bibitem{DiSAC}
P.~Christodoulou, ``{Soft actor-critic for discrete action settings},''
  \emph{arXiv preprint arXiv:1910.07207}, 2019.

\bibitem{DDPG}
T.~Lillicrap and et~al., ``{Continuous control with deep reinforcement
  learning},'' \emph{ICLR}, 2016.

\bibitem{cont4}
T.~Haarnoja, A.~Zhou, P.~Abbeel, and S.~Levine, ``{Soft actor-critic:
  Off-policy maximum entropy deep reinforcement learning with a stochastic
  actor},'' \emph{arXiv preprint arXiv:1801.01290}, 2018.

\bibitem{cont3}
D.~Silver and et~al., ``{Deterministic policy gradient algorithms},''
  \emph{ICML}, 2014.

\bibitem{Globe_far}
F.~Rezazadeh, H.~Chergui, L.~Alonso, and C.~Verikoukis, ``{Continuous
  Multi-objective Zero-touch Network Slicing via Twin Delayed DDPG and OpenAI
  Gym},'' in \emph{Proc. IEEE Glob. Commun. Conf.}, Dec. 2020.

\end{thebibliography}
\end{document}